# Meta-brain Models: biologically-inspired cognitive agents

Bradly Alicea[1,2] and Jesse Parent[1]

## Abstract

Artificial Intelligence (AI) systems based solely on neural networks or symbolic computation present a representational complexity challenge. While minimal representations can produce behavioral outputs like locomotion or simple decision-making, more elaborate internal representations might offer a richer variety of behaviors. We propose that these issues can be addressed with a computational approach we call meta-brain models. Meta-brain models are embodied hybrid models that include layered components featuring varying degrees of representational complexity. We will propose combinations of layers composed using specialized types of models. Rather than using a generic black box approach to unify each component, this relationship mimics systems like the neocortical-thalamic system relationship of the mammalian brain, which utilizes both feedforward and feedback connectivity to facilitate functional communication. Importantly, the relationship between layers can be made anatomically explicit. This allows for structural specificity that can be incorporated into the model's function in interesting ways. We will propose several types of layers that might be functionally integrated into agents that perform unique types of tasks, from agents that simultaneously perform morphogenesis and perception, to agents that undergo morphogenesis and the acquisition of conceptual representations simultaneously. Our approach to meta-brain models involves creating models with different degrees of representational complexity, creating a layered meta-architecture that mimics the structural and functional heterogeneity of biological brains, and an input/output methodology flexible enough to accommodate cognitive functions, social interactions, and adaptive behaviors more generally. We will conclude by proposing next steps in the development of this flexible and open-source approach.

## Introduction

The history of Artificial Intelligence (AI) has been marked by the rise of dominant models that have been hyped as being sufficient for simulating general intelligence. The hype surrounding such models has subsequently led to the exposure of flaws in both symbolic and connectionist models, and perhaps even the idea of general intelligence itself. Perhaps a few sources of inspiration can be found through embodied behavior, biological inspiration, and the strategic application of knowledge representation. The goal of replicating the rich behavioral repertoires and flexible internal models found in biological brains has consequences for constructing more realistic and informative models of intelligence. While it is obvious that biological nervous systems must be both heterogeneous and embodied, the correct combination of these attributes is ripe for innovation. In terms of emulation of biological intelligence in a computational system, this suggests that both subsymbolic (similar to what we will refer to as representation-free) and symbolic (similar to what we will refer to as representation-rich) are crucial components. This is where meta-brain models come into play: meta-brain models enable neuro symbolic models (Garcez and Lamb, 2020) to be expressed, refined, and extended in the form of embodied hybrid models.

---

[1] Orthogonal Research and Education Laboratory, Champaign-Urbana, IL
bradly.alicea@outlook.com
[2] OpenWorm Foundation, Boston, MA.

When we speak of representation, it refers to the interpretation of information in a manner that is decoupled from direct perception. A representation results in an internal mental state. An agent's internal state should reflect distinctive psychological functions located in different parts of a neural representation (Arbib, 1972). This should resemble the functional segregation of a biological brain, which is not a completely separable phenomenon (Deco et.al, 2015). The different layers of a meta-brain model thus consist of several interacting internal states that must be integrated. Representation can also be defined as the ability of one state to be transformed into something else (Newell, 1980). While this has interesting applications with respect to lower degrees of representation, richer levels of representation allow for a variety of functional attributes such as causality, intentionality, and object properties (Ramsey, 2007). As with biological brains, the meta-brain is not aware of how things in the world work, only that some attempts at solving problems work better than others (Dretske, 1988). We must also be careful to not ascribe too much intentionality to different levels of representation. The role of even the richest representation is not to provide conscious awareness or purpose, but to provide a joint probability map of stimulus and response (Eliasmith, 2005).

Our definition of representation is flexible, but can be defined in two ways. The first type of representation is a top-down representation, an example of which involves models organized around classificatory structures or symbolic frameworks. Specific models range from labeled input data to grounded symbols. This can be contrasted with a bottom-up representation, which involves topological structure that provides information about the sensory world or environment. Bottom-up systems range from cells that encode place/sensory fields to agent-based models with cells representing units of analysis of varying complexity.

**Taking an Embodied and Developmental Perspective**

Meta-brains draw from several diverse areas of inquiry such as AI, Neuroscience, Cognitive Science, Developmental Science, Cybernetics, and Cell and Molecular Biology. These fields provide a unique perspective on the core concepts most relevant to meta-brains: embodiment, developmental science, constructivism, and connectionism. Taken together, these features serve to extend meta-brain models from a simple case of a single sensory input and two layers to complex models with multiple layers, inputs, and even behavioral outputs. Throughout our discussion, we will expand upon our basic framework of layered, representational computational structures with features that make them relevant to developmental embodied agents, and eventually, to application in the broader world.

The most influential contributor to meta-brain model architectures involves embodiment. This includes both cognitive and Meta-brain models that are most effective when embodied in an agent with a body. Yet even in the disembodied case (Figure 1), the arrangement of layers and the sensory apparatus are embodied in a spatial context. Embodiment enables the information processing power of effectors, affordances, and various interfaces with the environment. Meta-brain models can also utilize the body of an agent as a mechanism for preprocessing sensory information, which would lie outside of the layered model structure. Morphological preprocessing would serve as a means to sample or filter information en route to the representation-free layer of the model.

In terms of developmental science, we can take advantage of two features of developmental agents: the generative nature of morphogenesis and compositional nature of learning. Our multilayered meta-brains and associated agent phenotypes rely upon morphogenetic growth, particularly innate rules that put these features of the model in place through a developmental process. Our conception of innateness is not simply to dictate what components a model should begin its life with, but also how it is shaped by interactions with the environment. As a complementary process, composition takes a number of forms in a meta-brain model, from the composition of a phenotype to the composition of complex behavioral outputs. Once the morphogenetic component of development is largely finished, a phenomenon called developmental freedom allows for an adaptive output to emerge from what becomes a largely static neural and morphological phenotype (Alicea et.al, 2020). As the developmental process stabilizes, the ability to construct coherent outputs and the informational synchrony of a meta-brain increases.

**Learning and Layering**

Drawing from learning itself, meta-brain agents take an unorthodox view to a traditional constructivist approach. In line with the layering principle of representational components, meta-brain models feature an innate component that provides a baseline as well as a heritability component to computational models of the brain. As a basement layer, the innate component resembles the genome in biological agents. In the spirit of creating efficient abstractions, innateness in meta-brain models consist of a transcription layer and a mutational layer. This is similar to mechanisms underlying epigenetic landscapes (Waddington 1962; Goldberg et.al, 2007; Bhattacharya et.al, 2011). The mutational layer is similar to a genetic encoding in that it consists of binary strings encoded with basic instructions on how to construct the brain and body of the computational agent. But the real power is in the mutation (and preservation) of this encoding. While mutations can introduce variation on a theme, our transcription layer allows for the mutational layer to be expressed in an analogue fashion. The transcription and mutational layers are generative in different ways, and provide a good balance of instructional information and response to particular environmental contexts.

The lower layer(s) of a meta-brain model are considered to be representation-free, the most common and powerful of which is the connectionist model. In their most basic implementation, connectionist models are generally representation-poor, and in their basic form homogeneous in structure (Katyal et.al, 2021). In meta-brain models, connectionist models can be heterogeneous in their structure, but crucially have connections to more representation-rich model types.

**Layering in Biological Brains**

In biological brains, laminar organization can be found in a number of contexts, with the greatest amount of characterization in the neocortex. Felleman and van Essen (1991) proposed a cortical hierarchy for understanding layering in the visual cortex of Macaque. In an object lesson for computational modeling, these layers do not have sharp functional or structural boundaries. Nevertheless, laminar structure is enforced by cytophysiological constraints which act to selectively constrain neural connectivity and activity (Miller, 1996). Such models have been refined to construct such as the laminar-based neural unit (Corbitt et.al, 2018), which characterizes the many neuronal cell types found in distinct cortical lamina. Using laminar-specific neuroimaging, we can better understand both the computational role of and connectivity between different cortical layers in cognitive functions such as language processing (Sharoh et.al, 2019), working memory (Finn et.al, 2019), and predictive processing (Haarsma et.al, 2020). Brain imaging studies have also found that the normal formation of a six-layered cortex is disrupted during embryonic and postnatal development in disorders such as autism (Stoner et.al, 2014) and microcephaly (Badea et.al, 2021).

**Basic Architecture of a Meta-brain Model**

Meta-brain models are hybrid models consisting of computational layers inspired by laminar anatomical structures. Each layer is a different type of computational model, and is meant to mimic structural differences inherent in many nervous systems. From a computational representation standpoint, meta-brain models provide us with neuro-symbolic capabilities while also selectively determining how neural and symbolic components interact. For example, the representation-rich component can "encase" or be "layered" on top of the representation-free component. The ability to toggle between these layers might be useful in a dynamical sense, or might allow us to better understand how our own brain selectively integrates information during cognitive processing. We have developed an open-source repository where a sizable collection of agent-based models are suggested for incorporation into our various layers of the model (https://github.com/Orthogonal-Research-Lab/Meta-brain-Models).

In Figure 1, a layered meta-brain is shown that can be encased structurally within other layers. Encasing in biological systems serves a number of diverse functions. Axons are encased by myelin, bone is encased by muscle and fatty tissue, different parts of the brain are encased by others, and germ layers define the earliest functional divisions in development. Often, encasing serves a functional role between tissues of different types, from chemical signaling between cells of distinct tissue types to more direct chemical communication using synaptic connections. Functionally, encasing emphasizes boundaries between tissue types or subsystems. In the case of meta-brains, the encasing we are approximating the outcomes of differential tissue growth, activity-dependent target tissue plasticity (Streidter, 2006), and more general developmental processes. In meta-brains, encasing can play a role in providing multiple access points between specific spatial locations in each layer. Topographical organization, or spatially explicit representations and projections, is a common organizational principle of the nervous system (Patel et.al, 2014). Meta-brain models provide a means to simulate the structure of such systems, and perhaps leverage their functional advantages as well.

**Embodiment and Regulation of the Meta-brain**

While meta-brain models are flexible, all implementations should have a set of common principles that guide their application. One of these principles is *embodiment*, which allows us to place our hybrid models in a broader context. We can also couple embodiment and the practice of encasing to create a subsumption architecture (Brooks, 1990). Subsumption involves the bottom-up processing of sensory stimuli, and has been considered to be biologically-inspired. In a meta-brain model, subsumption occurs through the processing of sensory information at progressively richer levels of representation. Unlike past implementations, the coordination of layers allows for representations of different complexities to emerge from representation-poor layers.

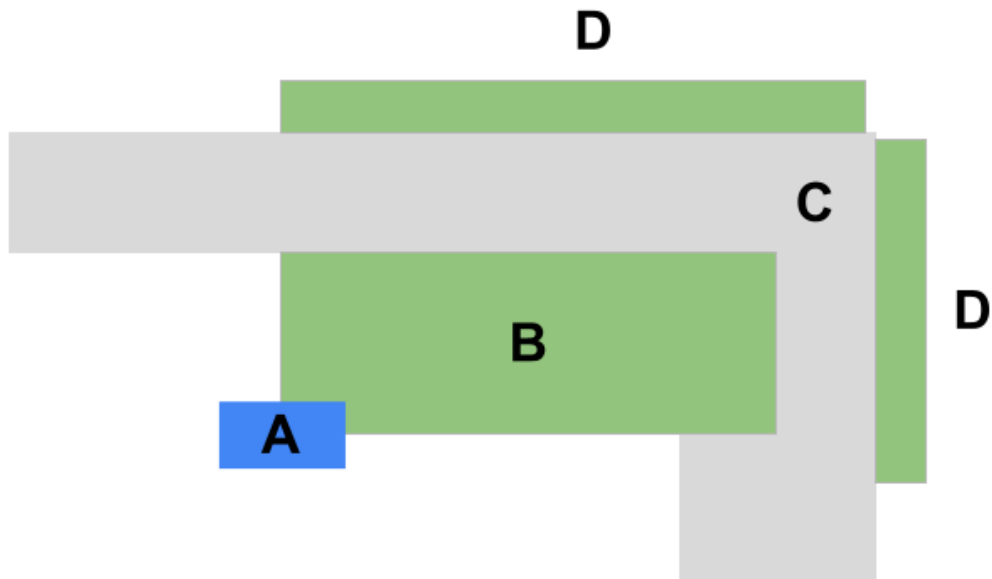

Figure 1. Structure of a meta-brain, encased in a geometry resembling a vertebrate brain. A: sensory input module. B: lowest representational layer, C: higher representational layer, D: highest representational layer.

**Cybernetic Regulation.** Subsumption architectures can be rather limiting when we consider meta-brains from the perspective of *cybernetic regulation.* While feedback is widely used in neuroscience at the level of systems such as the sensorimotor loop, feedback and other regulatory mechanisms are also critical to the communication and coordination between two meta-brain layers. In the language of meta-brain models, cybernetic control requires that the encasing layer (regulator) has more degrees of freedom than the underlying layer (controlled system). This results in a feedforward signal projecting towards higher degrees of representation, while feedback projects towards lesser degrees of representation. Aside from the potential for feedback between representational layers, interactions between layers necessitate one system to fully map to another. This requires a matching between the complexity of each layer. If a set of coupled models are vastly different in terms of their complexity (e.g. state space) then these deficiencies will lead to either insufficient control or severe errors in communication.

We also utilize the Every Good Regulator theorem (Conant and Ashby, 1970) to guide the composition of the various layer outputs and interactions. Meta-brains that conform to the EGRT allows for prediction and control given certain organizational principles. More generally, Ashby (1952) utilizes the homeostat to propose cybernetic control for brains as the natural selection of behavior patterns. Meta-brains are similar to a homeostat in that the connectivity between layers can both fluctuate and exhibit temporary interdependence.

Applying the cybernetics approach also suggests that meta-brains constitute a teleological regulatory system (Rosenblueth et.al, 1943) subject to the constraints of entropy and evolution. Although this is beyond the scope of our computational model, it is nevertheless an important principle to approximate. Connecting a meta-brain to cybernetic control also allows us to more rigorously analyze model outputs using the Free Energy Principle and related Bayesian approaches (Safron, 2021).

**Levels of Representation**

Meta-brain models address the issue of connectionist-symbolic integration by combining different degrees of representation into a single system. In the human brain, the representation of color relies upon a distributed set of layers that correspond to various features of the sensory phenomenon (Zeki and Marini, 1988; Seymour et.al, 2016). Meta-brains operate as a more general processing system which is not specialized for any specific property of the sensory world. A two-layer meta-brain might consist of a representation-free layer (connectionist model) and a representation-rich layer (object-oriented model). Meta-brains can also enable different levels of symbolism, from simple object classifications to cultural representations. In a multi-layer model, information processing can proceed orthogonally: information can be processed through the network or kernel of a single layer, while layer outputs can be utilized in a compositional manner to communicate between our layers of models. The relationship between layers can be defined as a homeomorphism, where the encasing layer is a richer model of the underlying layer.

**Isometric Integration.** The simplest way to join two or more layers of a meta-brain is through isometric integration. Previous work on utilizing isometric integration-like organization has been implemented in the form of layered control architectures (Prescott et.al, 1999; Wilson and Prescott, 2021). As shown in Figure 2, isometric layering can be demonstrated without incorporation in an anatomical or higher geometric setting. Nevertheless, isometrically arranged layers can exchange information in a geometrically-specific manner. This arrangement also demonstrates how layers can be combined to potentially leverage geometric constraints in a way that provides meta-structure to the interactions between layers. In Figure 2A, we present a three-layer laminar model, where the sparse representation receives inputs from both the representation-rich and representation-poor layers. A different configuration is shown in Figure 2B, where two sets of two-layer laminar are topologically distinct by connecting with a bidirectional connection. In this configuration, there is a duplicate sparse representation layer that serves to further coordinate each laminar set of layers.

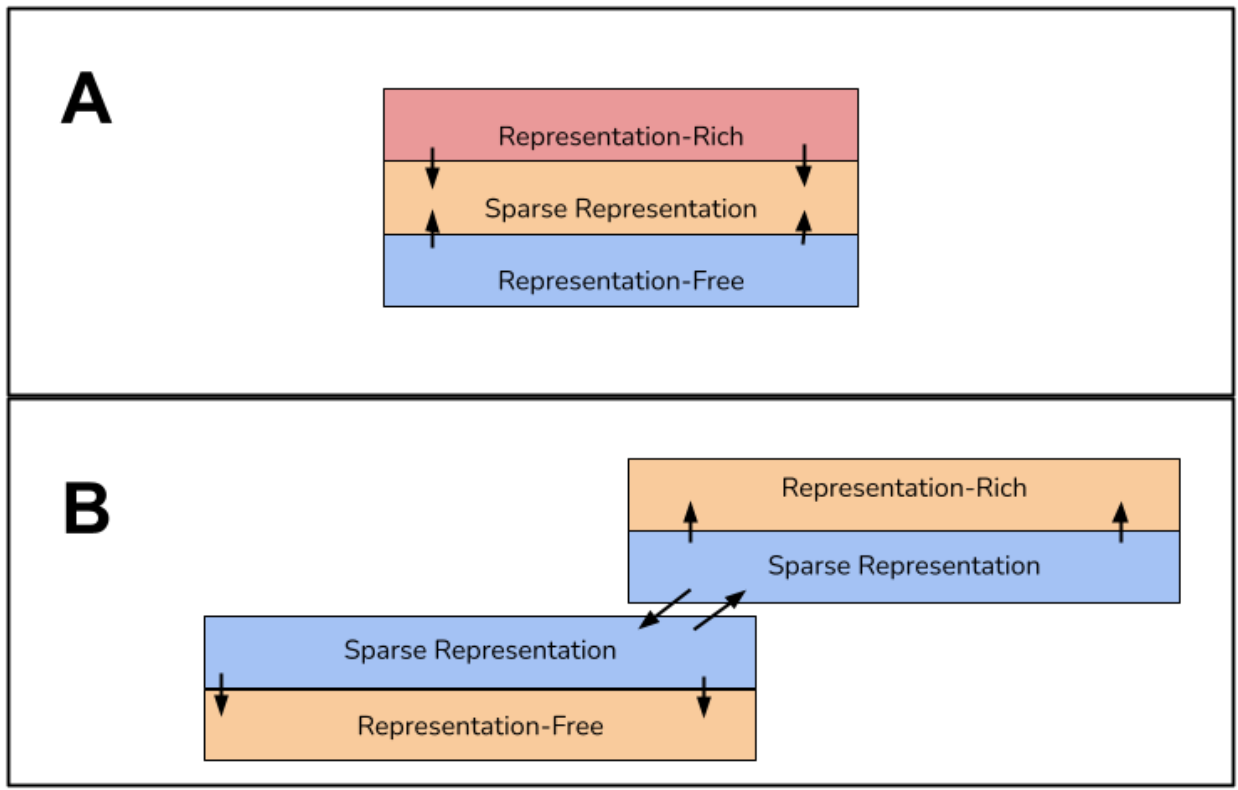

Figure 2. Basic isometric integration between the layers for two configurations of a meta-brain. A: a three-layer configuration with an arrangement that forms a representational gradient. B: an alternate meta-brain configuration with two sets of layers that form a spatially embodied and semi-discontinuous representational gradient.

**Embeddedness and Embodied Context**

Meta-brain models are incomplete without being embedded in an embodied context. By embodiment, we mean that an agent's physical body plays a significant role in shaping the cognitive capabilities of the agent. In Figures 1 and 2, meta-brain models are shown as disembodied entities. Yet embodiment can add adaptive capacity to an agent in addition to task performance utility (Bongard, 2010; Bredechel et.al, 2018). Meta-brains are thus best suited to being embedded in an anatomical context, represented by an agent with anatomical fidelity. Embodiment presumes that meta-brain will exhibit at least four properties: situatedness with respect to a particular context, time-dependence with respect to a task, the ability to offload computational requirements from the nervous system into the environment, and interactivity between a task and its environment (Wilson, 2002).

In Figure 1, we show a component called "sensory input". In the embodied context, this component becomes important as a mediator of direct perception (Gibson, 1979). Direct perception is a bottom-up approach to cognition that relies upon the environment to provide computational structure to the brain. In the case of radical embodied cognition (Chemero, 2009; Raja and Anderson, 2019), the case is made for a lack of computation in the brain itself, ascribing much of this capacity to direct perception. While the meta-brain perspective is not fully in accordance with this view, it does strike a balance between computationalism and non-computationalism. The lower layers of a meta-brain can lack representational capacity and be enriched by direct perception, which is in turn enabled by sensory inputs and the agent body. The higher layers of representational capacity then enable behaviors that are more intentional in nature.

This situation is ideal for things like direct learning (Jacobs and Michaels, 2007), which extracts a dynamical system from the environment presented to an agent. Enactive intelligence (DeLoor et.al, 2009; Froese and DiPaolo, 2011), or intelligence that arises through interactions between the agent's body and external environment, is also something that can be expanded upon using the meta-brain approach. In biological systems, morphology also plays a role in the development of mental representations. Body representations in the human brain act to create continuous feedback for motor control and positional awareness, as well as awareness of bodily position relative to spatial orientation (Carruthers, 2008). Meta-brain models do not utilize bodily representations *per se*, but such representations abstracted to computational models might be useful to integrate various aspects of the global context.

**Anatomical Fidelity.** One benefit of embodiment is the ability to incorporate the topology of anatomy into the computational model. With meta-brain models, there are two aspects of embodiment that provide anatomical fidelity to our models. The first benefit of embodiment is that the body itself engages in computation with respect to the brain-environment coupling. Even in agents, the effectors and their component parts can yield muscle synergies (Alessandro et.al,

2013) that regulate movement behavior in addition to cognitive offloading (Risko and Gilbert, 2016) capabilities. The second benefit, perhaps more generally useful to meta-brained agents, is that these encased and layered systems replicated the natural geometry of neural connectivity. Topological complexity in terms of adjacent layers provide benefits such as shorter wiring distances and tighter integration with an embodied agent's morphology.

Anatomical fidelity comes in the form of a layered morphology where some layers exhibit isometric integration and some act as parts of the morphology. However, they are all subject to developmental constraints under which layers grow according to a scaling law (West, 1999) and can also exhibit growth, movement, translation, and rotation (Thompson, 1917) with respect to other layers (Figure 3). Anatomical fidelity can also offer morphologically consistent affordances that enable a means for agents to change their size or shape in an adaptive manner with respect to behavior (Stokkermans et.al, 2021). Not only is this advantageous for the development of adaptive behaviors, but it also enables heterogeneity in the agent's structure and function.

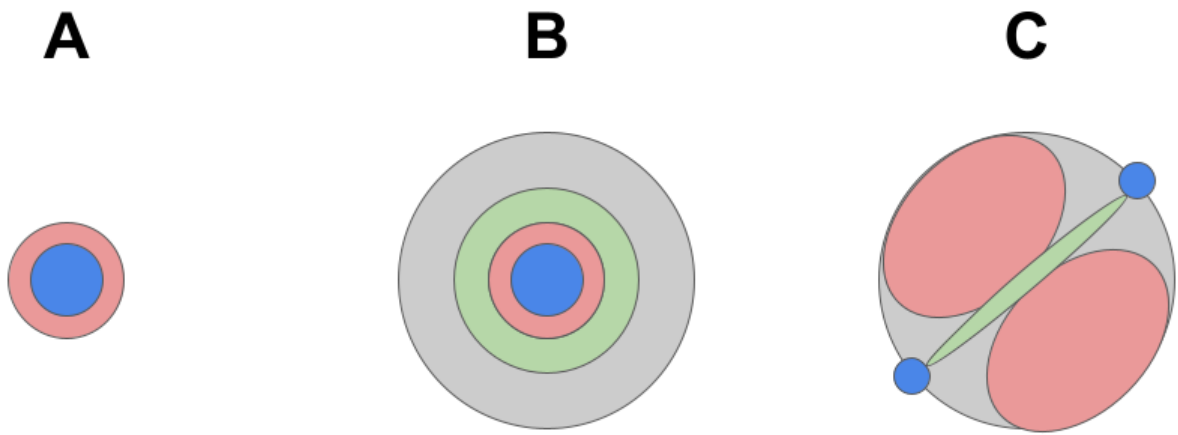

Figure 3. Example of anatomical fidelity as a developmental process. A: a two-layered embryonic agent. B: a four-layered developmental phenotype that expands in accordance with an 8/5 scaling law. C: four-layered mature phenotype with layers exhibiting growth, translation, movement, and rotation.

**Intrinsic and Extrinsic Factors: Heterogeneity**

To move beyond the input/output methodology without losing the benefits of computational and hypothetical simplicity, we can leverage the heterogeneity of a meta-brain model to provide a means for functional specialization. For example, heterogeneous meta-brains allow us to be explicit in terms of determining functional hierarchies like the kind we observe in the Primate visual system (Kruger et.al, 2013; Arcaro and Livingstone, 2017). We can also integrate and selectively turn off levels as needed for specific tasks. Heterogeneity also has benefits in terms of robustness and persistence of dynamical function. In general, diversity provides multiple paths to a specific function. Yet in the case of developmental meta-brain models, we can still preserve the ability to trace specific functional pathways. This type of architecture allows us to strike a balance between credit assignment (Minsky, 1961; Richards and Lilicrap, 2018) and emergent perception. Finally, single layers should exhibit temporal and topological heterogeneity to maximize representational stability and durability as discussed previously (Katyal et.al, 2021).

**Intrinsic and Extrinsic Factors: Innateness**

We can add a significant innate component to a layered model. Moving forward from a homogeneous model of an artificial nervous system, we need to enable dynamic attributes such as encodings and generativity. In extending meta-brain models to the Developmental AI paradigm, we will focus on evolution and development as means to imbue our models with innate mechanisms for generating intelligent behavior. One means to incorporate innateness is to provide a genetic representation for a given meta-brained agent. A genetic representation is a computational encoding that contains a generalized specification of an agent's body, brain, and behavior. While we do not propose using a typical genetic algorithm (GA), this encoding of the morphology and meta-brain layers should draw from the principles of GAs and genetic regulatory networks (GRNs). Using a GA/GRN architecture yields a set that can be gradually expressed throughout development and tuned to the agent's environment. Aside from selectively enabling plasticity, the GA/GRN architecture also imbues meta-brains with a basis for hereditary information as well as a source of developmental and functional invariance. In terms of our representational hierarchy, the genetic encoding is separated from the various levels of representation in the meta-brain itself, and serves as a generative system that provides a baseline of information to the agent without relying on the vagaries of environment or a more restrictive blueprint.

**Intrinsic and Extrinsic Factors: Development**

Meta-brains can also take inspiration from developmental (or ontogenetic) processes to provide an alternative to the so-called static adult mind (Oyama, 2000; Airenti, 2019). This provides a means to understand learning as both a compositional and morphogenetic process. One example of this is by using the example of small, emerging connectomes (Alicea, 2018). Development enables gradual differentiation of cells in the connectome, along with their formation of neural circuits. Meta-brains can approximate this process as simple connectomes at different times/stages of development serving as different layers and degrees of representations can model these processes coming online in an agent.

The development of meta-brains as emerging small connectomes can be described using the metaphor of Flatland. Flatland (Abbott, 1884) is a mythical world of only two spatial dimensions. The agents who inhabit Flatland do not have any concept of a third spatial dimension. And so it is with immature meta-brains, which develop their capabilities gradually in order of representational power. Only in the ontogenetic case it is representational capacity, not dimensionality, that defines the immature agent's perspectivist shortcomings. As development proceeds, representational capacity increases, and conceptual understanding of the third dimension emerges.

In terms of developing a form in which embodiment can take place, Morphogenesis might proceed by the innate component producing a precursor phenotype, which allows for the differentiation of multiple computational layers. As the layers differentiate, individual components of each layer come online. Then these layers need to be embodied in the agent's environment, which to an Flatland-dwelling agent resembles an unfamiliar world. In a simple meta-brain with only a single representation-free layer, the richness of the agent's environment is lost, depriving the developing agent of contextual information. As additional layers form that are more representationally robust, the environment can now be represented in a fuller form. However, a problem remains: the different layers of the meta-brain are at different developmental stages, and so are still unsynchronized in their ability to integrate information. This is where the interconnections between layers become important. These interconnections can be both feedforward and feedback in orientation, and work collectively to normalize differences in developmental stages.

**Intrinsic and Extrinsic Factors: Evolution**
Meta-brains can also evolve across generations of agents, and provides a means to think of meta-brain models from a dynamical, phylogenetic, and population perspective. This provides us with a means to explore the behavior of meta-brain models over long time intervals. The evolutionary trajectory of a given agent population is also influenced by its history of learning from sensory information (Hinton and Nowlan, 1987). By evolving the innate component of meta-brain models, we can create a phylogenetic representation of multiple generations. Given a mechanism for natural selection, this also allows us to produce different layers that are potentially adaptive to the problem at hand, and ultimately find the right degree of representation. The incorporation of natural selection and adaptation as a way to enable long-term meta-brain morphogenesis also requires populations of meta-brains, each exhibiting a range of variation. This variation could come in the form of network topologies at the single layer level, or as different patterns of interconnections between layers.

Finally, we can consider the developmental expression of various behaviors and layers from a *ontophylogenetic* approach (Cisek, 2019; Bennett, 2021). Cisek (2019) suggests that different parts of the brain expand in various taxonomic groups in the tree of life, which leaves a developmental signature consisting of areas that emerge early versus later in morphogenesis. A number of rudimentary behaviors shared by most species that have a brain are examined by Bennett (2021). This study reveals that many of these behaviors are implementable in agent-based systems such as Braitenberg Vehicles (Braitenberg, 1984) or other agents with minimal cognition (van Duijn et.al, 2006). This suggests selection may have favored an additive approach to building neurobehavioral complexity. In short, evolution is an important component of configuring meta-brain models.

## Application of Meta-brains to Specific Systems

The application of meta-brain models to specific systems also allows us to consider how developmental and evolutionary approaches might be used to build our layered heterogeneous representations in a specific environmental context. Of particular interest is the dynamic behavior of connections between each layer along with the establishment of new layers. The ability to toggle between layers, or the ability to use higher degrees of representation for some tasks and lower degrees of representation for others, might be valuable for a wide range of applications that require time-dependent processing or different levels of perspective for the same problem.

Aside from the further methodological refinement of meta-brains, there also exist a number of application domains that might be particularly amenable to this approach. These include tasks such as self-driving vehicles or agent navigation of unexplored environments. For these types of application, agents must employ a variety of behavioral and cognitive functions such as spatial cognition, learning, and phototaxis. Meta-brains might also be useful for more mundane tasks such as image processing, where segmentation and decomposition of an image can involve multiple levels of computational representation. To realize their full potential, the meta-brain approach also requires technical refinement on a number of fronts. One area that needs further technical detail is exploiting the connectivity patterns between layers. Another area to develop further is the topological nature of individual layers and the layering process more generally. A third domain for improvement is the sophistication of interactions with the environment via sensors and effectors.

### Meta-brain Implementation Contexts

Meta-brain agents can also be implemented in populations, where they experience interactions with other meta-brain agents. There are of course a number of coordination problems that become relevant. But social interactions also pose a host of unique sociocultural issues for computational agents (Bolotta and Dumas, 2022), including affordance learning and social embodiment. Affordance learning (Whiten, 2021) enables individual agents to learn higher-level representations from how other agents interact with their surroundings. In such cases, new meta-brain layers and their representational capacity can emerge soon after model initiation (during a developmental stage).

Social embodiment is also important for meta-brain development. From a representational standpoint, biological brains feature higher-level social representations that are scaffolded from lower-level sensorimotor representations (Meier et.al, 2012). Scaffolding (Wilson and Prescott, 2021) allows for many complex functions to emerge from brain lamina via functional constraints. In this way, meta-brains can feature representations of social relationships with select connections to sensory processing. Other forms of social behavior include swarm optimization and the emergence of collective behaviors (Ha and Tang, 2019; Gordon, 2019), which enable coordinated social behaviors between agents as well as the ability to learn these representations from one another.

Finally, the Developmental AI paradigm can be used to enable meta-brains to fully utilize the benefits of plasticity and constructivism that a set of developmental processes provides. Developmentally-inspired approaches such as Developmental Braitenberg Vehicles (Dvoretskii et.al, 2020), and overarching theories such as Continual Embodied Learning (Alicea et.al, 2020) all provide different means to incorporate features of developmental processes into a computational agent. One application domain for developmental embodied agents is spatial learning in small artificial nervous systems. Using a meta-brain to model spatial learning might not only enable behaviors such as path integration (Harootonian et.al, 2020), but also the acquisition of highly abstract spatial learning useful for navigating and analyzing high-dimensional, discontinuous topologies (Svarverud et.al, 2012).

**Future Directions**

To introduce the concept, we contextualize meta-brain models in terms of the history of Artificial Intelligence and inspiration from biological systems. Then we discuss the basic architecture of meta-brain models, including the three principles that guide implementation. This is followed by a discussion of architectural design features such as anatomical fidelity, heterogeneity, and cybernetic regulation. Meta-brain models also feature an innate component, which comes in the form of a base generative layer in addition to processes that interact with agent shape and structure. For example, agents are subject to processes such as development or evolution which affect form and function at multiple time scales. We then conclude with a discussion of applying meta-brains to specific systems, particularly with relevance to the Developmental AI approach.

While it is critical to understand the nature of connectivity between layers, and regulation between layers more specifically, we must also account for the independent representational drift of models that constitute different layers of a meta-brain. Representational drift (Clopath et.al, 2017; Rule et.al, 2019) occurs when changes to the environment or local content challenge the current state of the internal model. This has the potential to create a persistent source of error between layers (Rule et.al, 2019), which can interfere with their informational integration. Yet error can be controlled through the regulation of gain, as is the case in Xia et.al (2021). During viewing of a naturalistic stimulus that provides episodic activity, as long as gain can be dynamically regulated, the representation can remain stable. This is where the cybernetic approaches can assist us in building robust representations at multiple scales. More generally, internal representation has the potential to be seamlessly integrated with directly acquired sensory information. Through the measurement of mutual predictive information, predictions generated from an internal model and predictions extracted from environmental stimuli are not distinguishable (Candadai and Izquierdo, 2020).

**Broader Context of Intrinsic and Extrinsic Factors**

In terms of the interrelationship between morphogenesis and acquisition in development, developmental freedom is roughly analogous to neuroplasticity and connection weights in a representation-free model, but can also be captured using a form of reinforcement learning (RL). An RL perspective might be valuable at higher levels of representation, particularly in terms of overcoming the inconsistencies with plastic changes at lower representational layers. Similar to the concept of affordance learning, multi-agent RL (Canese et.al, 2021; Zhang et.al, 2019) would allow for meta-brain agents to acquire successful policies through the social learning of representations. Put into meta-brain terms, latent strategies (Xie et.al, 2020) are acquired from fellow agents in the form of representation-rich models that influence future policy selection. Returning to the problems of representational drift, it is also critical that the relationship between units within a layer are sufficiently different from other layers. In lesser representational layers, units should connect to other units using a series of transformational operations. Meanwhile, higher representational layers are connected through descriptive linkages (Barack and Krakauer, 2021).

Although our argument has relied upon the analogy between layers in an embodied brain and layering analogies in natural systems, it is not clear that biological levels are a coherent organizing principle in biology (Potochnik, 2021). Moreover, our division of representational levels is somewhat arbitrary, given that they rely on various types of computational models. Even the manner in which we define functional relevance and thus the relationships between levels have a much different cast when viewed from a phylogenetic perspective (Cisek, 2009). Another caveat we offer is that much of what we have presented has not yet been implemented or may not even be computationally rigorous. That said, we believe the component part described here can lead to models that improve upon current AI models, in addition to making such models much more biologically-inspired and robust.